\documentclass[pre,twocolumn,showpacs]{revtex4}
\usepackage{epsfig}
\usepackage{graphicx}

\begin{document}

\title{Anomalous Diffusion In Microrheology: A Comparative Study}

\author{I. Santamar\'{\i}a-Holek}
\affiliation{Facultad de Ciencias, Universidad Nacional Aut\'{o}noma de
M\'{e}xico. Circuito exterior de Ciudad Universitaria. 04510, D.
F., M\'{e}xico.}

\begin{abstract}
We present a comparative study on two theoretical descriptions of
microrheological experiments. Using a generalized Langevin
equation (GLE), we analyze the origin of the power-law behavior of
the main properties of a viscoelastic medium. Then, we discuss the
equivalence of the GLE with a generalized Fokker-Planck equation
(GFPE), and how more general GFPE's can be derived from a
thermo-kinetic formalism. These complementary theories lead to a
justification for the physical  nature of the Hurst exponent of
fractional kinetics. Theory is compared with experiments.
\end{abstract}
%\mtitle % DO NOT REMOVE

\maketitle

\section{Introduction}
In the last years, microrheology has become one of the most
important experimental techniques in soft condensed matter
\cite{trimper,mason2000,wirtz,finite}. It is a powerful technique
to determine the viscoelastic properties of complex fluids at time
and length scales complementary to those of classical rheological
methods \cite{mason2000,wirtz}. In these complex fluids, ranging
from polymer solutions and colloidal suspensions to the
intracellular medium, the presence of elastic forces and molecular
motors make anomalous the dynamics of the Brownian particles
\cite{finite,cromosomas,revMNET}. Hence, anomalous diffusion
becomes a central question that must be carefully analyzed due to
the presence of confinement and finite-size effects related with
particle dimensions. The dynamics of these particles can be
described through both, non-Markovian Langevin and Fokker-Planck
equations \cite{adelman, tokuyama,libroZ,donor,balescu}. Here, we
use and compare two different formalisms of anomalous diffusion
that in the linear force case are equivalent \cite{finite}.

The article is organized as follows. In Sec. {\bf II} we analyze
the generalized Langevin equation and its equivalence with a
non-Markovian Fokker-Planck equation. In Sec. {\bf III} we
introduce the thermokinetic description, show its equivalence with
the GLE formalism in the linear force case, and how it can be
generalized to the case of non-linear forces. Sec. {\bf IV} is
devoted to the conclusions.

\section{The Langevin approach to microrheology}

Consider a test Brownian particle of mass $m$ and radius $a$ with
position ${\bf x}(t)$ and velocity ${\bf u}(t)=d{\bf x}(t)/dt$.
The particle performs its Brownian motion through a viscoelastic
medium which can be made up by a solution of polymers or a
suspension of particles at sufficiently high concentration. In
these conditions, the motion of the particle could be restricted
to a small volume and, consequently, its dynamics may manifest
confinement and finite-size effects modifying the
anomalous behavior of the mean square displacement (MSD). This last
quantity is very important because the viscoelastic properties of
the medium can be inferred from it through the Stokes-Einstein relation \cite{finite}.

In a first approximation, the heat bath can be assimilated as an
effective medium that interacts with the particle by means of
elastic forces. In first approximation, these forces may be
represented through a harmonic force ${\bf F}_h=-\omega_m^2 {\bf
x}$, in which $\omega_m$ is a characteristic frequency
\cite{finite}. Then the dynamics of the particle can be described
by means of the generalized Langevin equation
\begin{equation}
\frac{d{\bf u}(t)}{dt}=-\omega_m^2 {\bf x}(t)- \int^{\infty}_0
\beta(t-\tau){\bf u}(\tau) d\tau+F(t), \label{gle-h}
\end{equation}
where $F(t)$ is a random force per unit mass and $\beta(t)$ is a
memory function called the friction kernel  \cite{adelman}. In the
Markovian case it takes the form: $\beta(t)=\beta_0
\delta(t-\tau)$ with $\beta_0=6\pi \eta a/m$ the Stokes friction
coefficient per mass unit, $\eta$ is the viscosity of the solvent and
$\delta(t-\tau)$ the Dirac delta function. Hence, in this case Eq.
(\ref{gle-h}) recovers its usual phenomenological form
\cite{libroZ}.

To describe the diffusion of the particle at sufficiently long
times, $t\gg\beta_0^{-1}$, two approximations can be followed.

\emph{a) The Overdamped Case.} In this case it is assumed that the
acceleration term of Eq. (\ref{gle-h}) can be neglected. As a
consequence the GLE takes the approximate form
\begin{equation}
\omega_m^2 {\bf x}(t)=- \int^{\infty}_0 \beta(t-\tau){\bf u}(\tau)
d\tau+F(t). \label{gle-overdamped}
\end{equation}
It is convenient to stress that in Eq. (\ref{gle-overdamped}),
the memory term involves the velocity of the particle.

\emph{b) Adiabatic Elimination.} In this second approximation, one
must first solve Eq. (\ref{gle-h}) in order to obtain
\begin{equation}
{\bf u}(t)=-\omega_m^2 \int^{\infty}_0 \chi(t-\tau) {\bf
x}(\tau)d\tau + F^*(t), \label{vel-gle-h}
\end{equation}
where we have used Laplace transforms, assumed the initial
velocity of the particle as equal to zero for simplicity, and
defined the memory function $\chi(t)$ as the inverse Laplace
transform $
%\begin{equation}
\chi(t)={\cal L}^{-1}\left\{[s+\tilde{\beta}(s)]^{-1}\right\}.
\label{chi}
%\end{equation}
$ This definition implies that $\chi(t)$ is the relaxation function
of the velocity of the particle \cite{adelman}. Moreover, we have
defined the time-scaled random force $F^*(t)$ by
\begin{equation}
F^*(t)=\int^{\infty}_0 \chi(t-\tau)F(t)d\tau .
\label{ruido-asterisco}
\end{equation}

Using now the identity ${\bf u}(t)=d{\bf x}(t)/dt=\dot{{\bf x}}$,
one then obtains the following generalized Langevin equation for
the position vector of the particle
\begin{equation}
\dot{{\bf x}}(t)=- \omega_m^2\int^{\infty}_0 \chi(t-\tau) {\bf
x}(\tau) d\tau+F^*(t). \label{gle-zwanzig}
\end{equation}

In the diffusion regime, $t\gg\beta_0^{-1}$, Eqs.
(\ref{gle-overdamped}) and (\ref{gle-zwanzig}) can be considered
as the GLE's for the position of the particle and then use them to
describe anomalous diffusion of a Brownian particle in a
viscoelastic heat bath. They constitute two different
models because Eq. (\ref{gle-zwanzig}) differs from
(\ref{gle-overdamped}) in the fact that it can be obtained through
an adiabatic elimination of variables, whereas the approximated
Eq. (\ref{gle-overdamped}) has been obtained by neglecting the
inertial term. Physically, the difference lies in the
fact that the relaxation function of the particle determining
the time dependence of the MSD, is different.

Quantitatively, these equations are not simply related one to each
other, since the memory kernel and the random force are not
equivalent due to the scaling introduced through $\chi(t)$. The
fluctuation-dissipation theorem (FDT) associated with Eq.
(\ref{gle-overdamped}) is
\begin{equation}
\langle F(t)F(0)\rangle=\beta(t), \label{fdt-over}
\end{equation}
where  the bracket $\langle \,\rangle$ means the average over
noise realizations. The corresponding FDT for the random force in
Eq. (\ref{gle-zwanzig}) is given by the expression
%\begin{equation}
$\langle F^*(t)F^*(t')\rangle=\int^t_0 d\tau \int^{t'}_0 d\tau'
\chi(t-\tau)\chi(t'-\tau') \beta(\tau-\tau')$, \cite{libroZ}.
%\label{fdt-zwanzig}
%\end{equation}
%where we have used (\ref{fdt-over}). In general, Eqs.
%(\ref{fdt-over}) and (\ref{fdt-zwanzig}) are not simply related.
However, in the case when the friction kernel is proportional to a
Dirac delta function, Eq. (\ref{fdt-over}) defines a thermal
(white) noise. After performing the adiabatic elimination of
variables, one obtains the relation $\langle F^*(t)F^*(0)\rangle
\sim exp(-\beta_0 t)$ for the time-scaled random force. This
implies that the position ${\bf x}(t)$ of the particles satisfies
a non-Markovian equation with a exponentially decaying time
correlation of the random force, \cite{libroZ}. Only in the
Markovian case Eqs. (\ref{gle-overdamped}) and (\ref{gle-zwanzig})
are equal.

At certain time scales, anomalous diffusion is characterized by a
power-law behavior of the MSD of the particle as a function of
time. Within a Langevin description, this behavior is a
consequence of the statistical properties of the random force,
that is, of the heat bath as follows from the relations already
obtained.
In particular, by using Eqs. (\ref{gle-overdamped}) and
(\ref{fdt-over}), it can be shown that anomalous diffusion is a
consequence of a fractionary Gaussian noise (FGN) that satisfies
the FDT \cite{donor}
\begin{equation}
\langle F(t)F(0)\rangle=({k_BT}/{m})\beta_0^2
2H(2H-1)|\tilde{t}|^{2H-2}, \label{fdt-FGN}
\end{equation}
where the exponent must satisfy the
relation: $1/2\leq H\leq1$, the factor $\beta_0^2$ gives
the correct dimensions and we
have introduced the dimensionless time $\tilde{t}=t_0^{-1} t$,
with $t_0$ a characteristic time. In particular,
$t_0$ can be taken as the characteristic
diffusion time $\tau_D=a^2D_0^{-1}$, with $D_0=k_BT/(m \beta_0)$.
Hence, using Eqs.
(\ref{gle-overdamped}) and (\ref{fdt-FGN}) one may calculate the
time dependence of the MSD by using the method of Laplace
transforms. Here, it is convenient to write (\ref{gle-overdamped})
in the dimensionless form
\begin{equation}
\tilde{{\bf x}} (\tilde{t})=- \alpha \int^{\infty}_0
\tilde{\beta}(\tilde{t}-\tilde{\tau})\tilde{{\bf u}}(\tilde{\tau})
d\tilde{\tau}+\tilde{F}(\tilde{t}). \label{gle-over-ad}
\end{equation}
where $\tilde{\beta}(\tilde{t})=|\tilde{t}|^{2H-2}$ and
$\tilde{F}=a^{-1}\omega_m^{-2}F$ are dimensionless quantities  in
accordance with Eqs. (\ref{fdt-FGN}) and (\ref{fdt-over}), and
$\alpha=2H(2H-1)\beta_0^2\omega_m^{-2}$. In writing
(\ref{gle-over-ad}) we have used $\tilde{t}$ and the dimensionless
position vector $\tilde{{\bf x}}=a^{-1} {\bf x}$.

Now, by Laplace transforming Eq. (\ref{gle-over-ad}), one may
obtain the expression for the transformed velocity $\hat{{\bf
u}}(s)$, which in turn can be substituted into the definition
$s\hat{{\bf x}}(s)=\hat{{\bf u}}(s)$ with ${\bf x}(0)=0$ for
simplicity. After solving for $\hat{{\bf x}}(s)$ one obtains
\begin{equation}
\hat{{\bf x}}(s) =
\frac{\alpha^{-1}\Gamma^{-1}(2H-1)s^{2H-2}}{1+\alpha^{-1}
\Gamma^{-1}(2H-1) s^{2H-2}} \hat{F}(s). \label{x(s)}
\end{equation}
where $\Gamma(2H-2)s^{1-2H}$ and $\hat{F}(s)$ are the Laplace
transforms of $\tilde{\beta}$ and $\tilde{F}$, respectively. The
inverse Laplace transform of Eq. (\ref{x(s)}) is given in terms of
the convolution of the noise $\tilde{F}$ and the Mittag-Leffler
polynomial $z_*^{1-2H} E_{2-2H,2-2H}\left(- z_*^{2-2H}\right)$,
with the scaled variable $z_*=\left[\alpha
\Gamma(2H-1)\right]^{-1/(2-2H)}\tilde{z}$, \cite{saxena}.

The expression for the MSD can now be obtained by
calculating $\langle x^2\rangle(t)$ in terms of
the inverse Laplace transform of Eq. (\ref{x(s)}),
using the FDT (\ref{fdt-FGN}) and
performing the corresponding integrals. The general
expression involving Mittag-Leffler polynomials is complicated,
however at
times $t$ satisfying $\beta_0^{-1}\ll t \sim t_0$,
one obtains the following short-time power law behavior
\begin{equation}
\langle x^2\rangle(t) \sim
\left({k_BT\beta_0^2}/{m\omega_m^4}\right)
\left({t}/{t_0}\right)^{2-2H},
 \label{MSD-over}
\end{equation}
where some constants have been dropped for the sake of simplicity.
%The procedure and assumptions we have used here, are similar to
%those of previous theoretical descriptions, see, for instance,
%Ref. \cite{mason2000}.

\begin{figure}[tbp]
%\begin{center}
\includegraphics[width=0.8\linewidth]{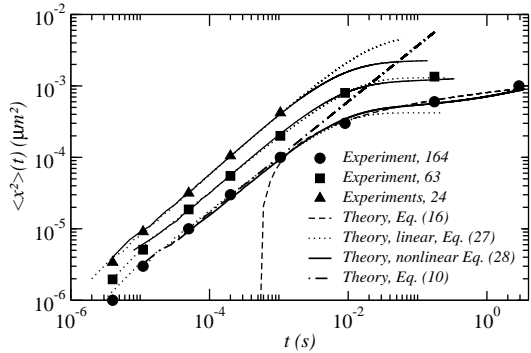}
%\end{center}
\caption{MSD vrs time. Comparison between theory (lines) and
experiments with actin filament networks (symbols) (taken from
Ref. \cite{wirtz}). The dashed line corresponds to the power law
behavior valid for long times, Eq. (\ref{MSD-Smol-Adi}) with
$t_0=2\cdot 10^{-3}s$, $\omega_m\sim 2.6\cdot10^{2}s^{-1}$, and
$b\sim 1\cdot10^{-2}$. The dotted lines correspond to a stretched
exponential behavior Eq. (\ref{xx-adim}), with $\omega
_{T}^{2}/\omega_{m}^{2}\sim5.61\cdot10^{-4}$, $2\omega _{m}^{2}B_2
t_0/\alpha \beta_0\sim35$ and ${2\beta_0}/{\alpha
t_0\omega_m^2}=3/4$. The solid lines correspond to the nonlinear
model Eq. (\ref{MSD-nonlinear}). The values of the parameters are
indicated in the text. The dash-dotted line corresponds to Eq.
(\ref{MSD-over}) with $H=5/8$. The numbers $24$, $63$ and $124$
correspond to increasing concentrations of actin filaments,
\cite{wirtz}. } \label{CompWirtz1}
\end{figure}

In Figure {\bf \ref{CompWirtz1}}, we compare the MSD given by Eq.
(\ref{MSD-over}) (dash-dotted lines) with experiments (symbols) of
$0.965 \mu m$ diameter latex microspheres imbedded in F-actin
solutions of increasing concentrations, \cite{wirtz}. As expected,
the agreement between theory and experiments is good for short
times $\beta_0^{-1}\ll t \sim t_0$ with $t_0\sim\tau_D$ playing
the role of a crossover time. If we take $\beta_0\sim
10^{6}s^{-1}$ then $\tau_D \sim 10^{-2}s$ in typical situations.

The MSD can be used to obtain the microrheological properties of
the medium by means of the generalized Stokes-Einstein relation
\cite{mason2000}. From Eq. (\ref{gle-h}) in the zero inertia
approximation, one may show that the complex shear modulus of the
medium $G''(\omega)$ satisfies the relation
\begin{equation}
G''(\omega) \simeq \left({k_BT}/{m \omega \langle
\hat{x}^2\rangle(\omega)}\right), \label{G}
\end{equation}
where $\omega$ is the frequency. Now, by taking the Fourier
transform of Eq. (\ref{MSD-over}) and substituting the result into
(\ref{G}), one obtains the scaling relation $G''(\omega) \sim (t_0
\omega)^{2-2H}$.

From this analysis it follows that the power-law behavior of the MSD
of the particles and the related complex shear modulus of the
effective medium, are a consequence of the fractionary
Gaussian noise of the GLE (\ref{gle-h}). Notice that
Eq. (\ref{gle-zwanzig}) can also be used to describe anomalous
diffusion, however this will be explained after Eq. (\ref{Smol-Osc}),
below.

%%%%%
\emph{The Generalized Fokker-Planck Equation.} An equivalent
approach to study anomalous diffusion in a viscoelastic bath can
be performed by means of a generalized Fokker-Planck equation .

By using the Laplace transform technique,  in Ref. \cite{adelman}
it has been shown that the GLE (\ref{gle-h})
is equivalent to the following non-Markovian
Fokker-Planck equation containing time dependent coefficients
\begin{eqnarray}
\frac{\partial }{\partial t}f+\frac{\partial f}{\partial {\bf x}}
\cdot \left( {\bf u}f\right)- \tilde{\omega}^2(t){\bf x}\cdot
\frac{\partial f}{\partial {\bf u}}= \psi(t)
\frac{\partial}{\partial {\bf u}}\cdot\frac{\partial f}{\partial
{\bf x}}
\nonumber \\ +\frac{\partial }{%
\partial {\bf u}} \cdot \tilde{\beta}(t) \left[{\bf u}f+ \frac{k_{B}T}{m}
\frac{\partial f}{\partial {\bf u}}\right] . \label{GFP}
\end{eqnarray}
Here, $f({\bf x},{\bf x}_0;{\bf u},{\bf u}_0,t)$ is the
probability distribution function depending on the instantaneous
position ${\bf r}$ and velocity ${\bf u}$ of the Brownian
particle. ${\bf r}_0$ and ${\bf u}_0$ are the corresponding
initial conditions. In this equation, memory effects are
incorporated through the time dependent coefficients
$\tilde{\beta}(t)$ and $\tilde{\omega}(t)$ and
$\psi(t)=({k_{B}T}/{m\omega_m^2})\left[\tilde{\omega}^2(t)-\omega_m^2\right]$. They are given in terms of combinations of the relaxation functions
$\chi_u(t)=m(3k_BT)^{-1}\langle{\bf x}(t)\cdot {\bf u}_0\rangle$
and $\chi_x(t)=\langle {\bf x}_0^2\rangle^{-1}\langle{\bf
x}(t)\cdot {\bf x}_0\rangle$, and their
%\begin{eqnarray}
%\tilde{\beta}=-d\ln|\Delta|/dt\,\,\,\text{and}\,\,\,
%\tilde{\omega}^2=\Delta^{-1}(\ddot{\chi}_u\dot{\chi}_x-\ddot{\chi}_x\dot{\c%hi}_u),
%\label{coefs(t)}
%\end{eqnarray}
explicit form is given in Ref. \cite{adelman}. It is worth
to stress here that an equation similar to (\ref{GFP}) has
also been obtained by using projector operator techniques in
\cite{tokuyama}.

The result expressed in Eq. (\ref{GFP}) is important, because it
implies that generalized Fokker-Planck equations incorporating
memory effects through memory functions constitute models not
simply related with Eq. (\ref{gle-h}), \cite{finite,trimper}.
Passing from one model to the other implies an approximation that
imposes conditions on the rate of change of the fields involved in
the description \cite{finite,libroZ,balescu}.

At times $t\gg\beta_0^{-1}$, from Eq. (\ref{GFP}) one may derive a
generalized Smoluchowski equation for the mass density $\rho({\bf
x},t)=m\int fd{\bf u}$. To derive it, one may follow the adiabatic
elimination of variables procedure by calculating the evolution
equations for the first three moments of $f({\bf x},{\bf u},t)$.
Then, after imposing the condition of large times
$t\gg\beta_0^{-1}$, a reduced description is obtained in terms of
an evolution equation for $\rho$.

Averaging Eq. (\ref{GFP}) over ${\bf u}$ assuming that the
currents vanish at the boundaries, we obtain the continuity
equation
\begin{equation}
{\partial \rho }/{\partial t}=-\nabla\cdot (\rho {\bf v}),
\label{continuity}
\end{equation}
where ${\bf v}({\bf x},t)=m\rho^{-1}\int {\bf u}fd{\bf u}$ is the
average velocity field.

At long times one obtains the following constitutive relation for
the diffusion current \cite{PHYSA03}
\begin{eqnarray}
\rho {\bf v} \simeq - \tilde{\beta}^{-1} \tilde{\omega}^2 {\bf x}
\rho- \tilde{D}\nabla \rho , \label{dif-current}
\end{eqnarray}
where we have defined the effective diffusion coefficient
$\tilde{D}(t)=({k_BT}/{m\omega_m^2})\tilde{\beta}^{-1}\tilde{\omega}^2$
and used the definition of $\psi(t)$. Substitution of
Eq. (\ref{dif-current}) into (\ref{continuity}) leads to
\begin{equation}
{\partial \rho }/{\partial t}=\tilde{D}(t)\left[\nabla ^{2}\rho -
\left({m\omega_m^2}/{k_BT}\right)\nabla \cdot \left( \rho {\bf
x}\right)\right], \label{Smol-Osc}
\end{equation}%
which constitutes the generalized Smoluchowski equation (GSE) for
$\rho$. An equation similar to (\ref{Smol-Osc}) can also be
obtained from Eq. (\ref{gle-h}) by using Laplace transforms. In
Ref. \cite{oxtoby}, the authors found that
$({m}/{k_BT})\tilde{D}(t)$ is proportional to $ - d \ln
\left|R(t)\right|/dt$, with $R(t)=\chi(t)/\chi(t_0)$ and
$\chi(t)=\langle {\bf x}\cdot{\bf x}_0\rangle(t)$ the relaxation
function of the position of the particle. If one considers the
case of FGN, the relaxation function $R(t)$ is given in terms of a
Mittag-Leffler polynomial leading to a complicated form for the
effective diffusion coefficient  $\tilde{D}(t)$, \cite{donor}. By
evaluating numerically the obtained relation, it can be shown that
at short times the function follows a power law behavior with an
exponent equal to that calculated by simply taking the time
derivative of Eq. (\ref{MSD-over}). At larger times, a crossover
arises in which  the relaxation function can be assumed to decay
as the power law $R(\tilde{t})\sim \tilde{t}^{-b}$. Using the
definition $\langle x^{2}\rangle=\int x^{2}\rho \,d{\bf x}$, Eq.
(\ref{Smol-Osc}) and assuming $\tilde{\beta}^{-1}\tilde{\omega}^2=
\beta_0^{-1} \omega^2_m \tilde{\gamma}(\tilde{t})$, with
$\tilde{\gamma}$ dimensionless, one may calculate the following
expression for the MSD
\begin{equation}
\langle x^{2}\rangle\sim
3({k_BT}/{m\omega^2_m})\left[1-\left({t}/{t_0}\right)^{-2b}\right].
\label{MSD-Smol-Adi}
\end{equation}
At times $t\gg t_0$, Eq. (\ref{MSD-Smol-Adi}) can be used to fit
the experimental results, as shown by the dashed line in Figure
{\bf \ref{CompWirtz1}}. See the discussion after Eq.
(\ref{MSD-over}).

This analysis shows that, in the linear force case the GLE with
FGN offers a good explanation of the subdiffusion observed in
microrheological experiments. A different approach in
order to have deeper understanding of the physical mechanisms
responsible for subdiffusion in a viscoelastic fluid will be presented in the following section.

\section{Thermokinetic-description of anomalous diffusion}

The Fokker-Planck type Eqs. (\ref{GFP}) and (\ref{Smol-Osc}) for
the probability distribution function, admit a different
interpretation in terms of kinetic equations. This
reinterpretation enriches and simplifies the description of
anomalous diffusion, because allows one to formulate
phenomenological models for the coefficients $\tilde{\beta}$ and
$\tilde{\omega}$ which not do depend on the \emph{a priori}
election of the statistical properties of the random force of Eq.
(\ref{gle-h}).

Fokker-Planck type kinetic equations for different physical
phenomena can be obtained by using two general principles. The
first one is the conservation of the probability, expressed by the
equation
\begin{equation}
{\partial f({\underline{\Gamma}},t)}/{\partial
t}=-\nabla_{\Gamma}\cdot \left[f(\underline{\Gamma},t) {\bf
V_{\gamma}}\right], \label{cont-prob}
\end{equation}
where $\underline{\Gamma}$ represents the set of variables
necessary to describe the state of the system at a mesoscopic level.
Here, $f(\underline{\Gamma},t) {\bf V_{\gamma}}$ represents a
diffusion current in $\underline{\Gamma}$-space and
$\nabla_{\Gamma}$ the corresponding gradient operator. The second
principle is the so-called generalized Gibbs entropy postulate
\cite{revMNET,groot}
\begin{equation}
\rho \Delta s (t)= -k_B \int f(\underline{\Gamma},t)
\ln\left|{f(\underline{\Gamma},t)}/{f_{leq}(\underline{\Gamma})}
\right|d\underline{\Gamma}, \label{Gibbs}
\end{equation}
where $\Delta s$ is the difference of the specific entropy with
respect to a local equilibrium reference state, characterized
through the probability distribution
$f_{leq}(\underline{\Gamma})$. The assumption of local equilibrium
allows one to calculate $f_{leq}(\underline{\Gamma})$ by using the
techniques of equilibrium statistical mechanics.
%Eq. (\ref{Gibbs})
%constitutes an irreversibility criterion that guaranties that the
%obtained equations satisfy an extremal principle in
%$\Gamma$-space.

To obtain the evolution equation for $f(\underline{\Gamma},t)$, we
may follow the rules of mesoscopic nonequilibrium thermodynamics
(MNET) \cite{revMNET,Adam07}. Schematically, by taking the time
derivative of Eq. (\ref{Gibbs}) and inserting (\ref{cont-prob}),
after integrating by parts assuming the usual boundary conditions,
one obtains the entropy production $\sigma(t)$
\begin{equation}
\sigma(t)=-\frac{m}{T} \int f(\underline{\Gamma},t) {\bf
V_{\gamma}} \cdot \nabla_{\Gamma} \left[ \frac{k_BT}{m}
\ln\left|{f}/{f_{leq}}\right|\right]d\underline{\Gamma}.
\label{sigma}
\end{equation}
In a similar way as in linear irreversible thermodynamics
\cite{groot}, we can assume linear relationships between the
forces $({k_BT}/{m})
\ln\left|{f}/{f_{leq}(\underline{\Gamma})}\right|$ and the
conjugated currents $f {\bf V_{\gamma}}$. Using them, one finally
arrives at the multivariate Fokker-Planck equation
\begin{equation}
{\partial f}/{\partial t}=\nabla_{\Gamma} \cdot
\underline{\underline{\tilde{\xi}}}(t)\left[({k_BT}/{m})
\nabla_{\Gamma} f - \left( f {\bf X}\right)\right],
\label{F-P-MENT}
\end{equation}%
where the generalized force ${\bf X}(\underline{\Gamma})$ is
defined in the usual form: ${\bf X}=-{k_BT}\nabla_{\Gamma} \ln
f_{leq} $, \cite{libroZ}. Here, memory effects are incorporated
through the time dependent Onsager coefficients introduced by the
linear coupling between forces and fluxes, and contained in
$\underline{\underline{\tilde{\xi}}}(t)$, \cite{finite,PHYSA03}.
The tensorial character of $\underline{\underline{\tilde{\xi}}}(t)$
accounts for the anisotropy of the medium.

If we consider a Brownian particle under the action of a linear
force and whose mesoscopic state is determined by its
instantaneous velocity ${\bf u}$ and position ${\bf x}$, Eq.
(\ref{F-P-MENT}) takes the form (\ref{GFP}). In this case, the
local equilibrium distribution is
$f_{leq}=f_0exp\left[-(k_BT/2m)\left({\bf
u}^2+2\phi\right)\right]$, where $\phi({\bf x})$ is the potential
associated with the harmonic force, $f_0$ is a normalization
factor and we have assumed that in $\Gamma$-space entropic forces
are coupled to the currents independently from those arising from
an energy potential, reflecting its different physical origin
\cite{PHYSA03}.

At larger times, we may assume that the state of the particle is only
determined by its position vector, then the local equilibrium
distribution is
$\rho_{leq}=\rho_0exp\left[-(k_BT/m)\phi({\bf x})\right]$, where
the potential $\phi({\bf x})$ is arbitrary.  By following the
procedure indicated above, the following GSE can be derived
\begin{equation}
{\partial \rho }/{\partial t}=D(t)\left\{\nabla ^{2}\rho -
\left({m}/{k_{B}T}\right)\nabla \cdot \left[ \rho {\bf X}({\bf
x})\right] \right\},  \label{Smol-MNET}
\end{equation}%
where we have assumed an isotropic medium when writing the scalar
effective diffusion coefficient $D(t)$ and defined ${\bf
X}=-\nabla \phi({\bf x})$. Hence, the thermokinetic formalism
allows one to derive generalized Fokker-Planck  equations of the
form (\ref{Smol-MNET}) which contain nonlinear forces. No
assumptions on the statistical nature of the random force have
been done until now. This is an important difference with respect
to the GLE description, because the method presented in the second
section gives exact Fokker-Planck equations only in the case of
linear forces.

During its motion through the viscoelastic medium, the Brownian
particle induces perturbations on the velocity field of the host
fluid surrounding it. These perturbations propagate and are
reflected by the local boundaries which can be made up by a
polymer network or another suspended particles. As a consequence,
they modify the velocity field of the fluid around the particle in
a later time, introducing memory effects. In this form,
hydrodynamic interactions change the distribution of stresses used
to calculate the force over the surface of the particle. In first
approximation, this force is characterized by the effective
mobility coefficient $\gamma (t)=(m/k_BT)D(t)$, in the form
\cite{finite}
\begin{equation}
\gamma (t)=\beta _{0}^{-1}\alpha \tilde{\gamma}(t).  \label{gamma-finite}
\end{equation}%
Here, $\tilde{\gamma}(t)$ is a dimensionless function of time and
$\alpha =1+B_{1}{a}/y$ is the mentioned correction
\cite{finite,brenner,saarlos}. The coefficient $B_1$ depends on
the nature of the boundary (solid wall, membrane, polymer network
or a cage formed by surrounding particles) and $y$ represents a
characteristic length of the medium.

By using the dimensionless variables previously introduced and
defining the time scaling $\tau(t)=\int \tilde{\gamma}(t)dt$, from
(\ref{Smol-MNET}) one obtains
\begin{equation}
{\partial \rho }/{\partial \tau}= \alpha \left({\omega
_{T}^{2}}t_0/{\beta
_{0}}\right)\tilde{\nabla}^{2}\rho -\alpha%
\left({F_{0}t_0}/{\beta_{0}a}\right) \tilde{\nabla}%
\cdot \left( \rho \tilde{{\bf X}}\right) ,  \label{dif-harmon}
\end{equation}%
where $\omega _{T}^{2}\equiv k_{B}T/ma^{2}$ and $F_{0}$ is the
magnitude of the force per unit mass ${\bf X}$ due, for instance,
to the polymer network.

The time dependence of $\tau$ can be
obtained by using the
evolution equation for the time correlation function
$\chi(\tilde{t})=\langle \tilde{{\bf x}}\cdot\tilde{{\bf x}}_0\rangle(\tilde{t})$
(see, Eq. {\ref{Smol-Osc}}), \cite{finite}
\begin{equation}
{d \chi(\tilde{t})}/{d \tilde{t}} = -\alpha
\left({F_{0}t_0}/{\beta_{0}a}\right) \tilde{\gamma}(\tilde{t})
\langle \tilde{{\bf X}}\cdot\tilde{{\bf x}}_0\rangle(\tilde{t}).
\label{evol-Xi}
\end{equation}
Eq. (\ref{evol-Xi}) involves a complete hierarchy for the moments
of the distribution $\rho$. However, in the linear force case it
takes a closed form leading to the relation
\begin{equation}
\gamma(\tilde{t}) =
d \ln \left|R^{-\beta_0/\alpha t_0 \omega_m^2}\right|/d\tilde{t},
\label{tau-Xi1}
\end{equation}
where we have used ${\bf X}=-\omega_m^2{\bf x}$ with
$\omega_m^2=a^{-1}F_0$ and $R(\tilde{t})=\chi(\tilde{t})/\chi(0)$.
Using the definition and Eq. (\ref{tau-Xi1}) one gets
$\tau(\tilde{t})=\ln \left|R^{-\beta_0/\alpha t_0\omega_m^2}\right|$. These
relations imply: $D=\tilde{D}$, see the discussion after Eq. (\ref{Smol-Osc}).
\begin{figure}[tbp]
%\begin{center}
\includegraphics[width=0.6\linewidth]{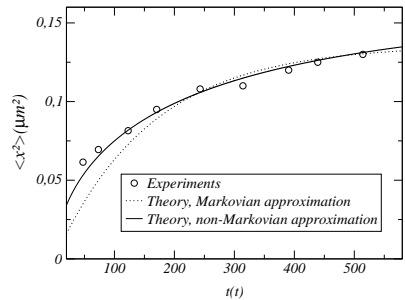}
%\end{center}
\caption{MSD vrs time. Comparison between theory (lines) and
experiments of chromatin diffusion in living cells.
The dotted line corresponds to a linear model in the Markovian
approximation. The data were taken from Ref. \cite{cromosomas}
(circles) and the figure from \cite{signpost}. }
\label{FigCromosomas}
\end{figure}

At short times, we can infer the time dependence of
$\tau(\tilde{t})$. Expanding the logarithm in its argument around
the unity: $\ln|
R^{-\,{\beta_0}/{\alpha t_0 \omega_m^2}}|\simeq
R^{-{\beta_0}/{\alpha t_0\omega_m^2}}-1 + O(R^2)$ and
taking into account that $R(\tilde{t})$ must be an even function
of time, a first order expansion leads to:
$R^{-1}(\tilde{t})\simeq 1+B_2^*\tilde{t}^2+O(\tilde{t}^4)$. Thus,
we obtain the approximate expression
\begin{equation}
\tau(\tilde{t}) \sim B_2
\tilde{t}^{\,{2\beta_0}/{\alpha t_0\omega_m^2}},
\label{tau-Xi}
\end{equation}
where the parameter $ B_2 \propto B_2^*$ in general depends on
the characteristic length of the medium \cite{finite}. Important to notice
is that the exponent depends on $\beta_0$,
$t_0$, $\omega_m$ and $\alpha$.  Eq. (\ref{tau-Xi}) is consistent
with a stretched exponential relaxation
of the correlations \cite{Agus05}. Hence, in the linear case,
Eq. (\ref{dif-harmon}) gives the MSD
\begin{equation}
\langle \tilde{x}^{2}\rangle (\tau )=3(\omega _{T}^{2}/\omega
_{m}^{2})\left[ 1-e^{-{2\alpha\omega _{m}^{2} t_0\tau(t) }/{%
 \beta_0}}\right] ,
\label{xx-adim}
\end{equation}%
which at short times yields
$\langle \tilde{x}^{2}\rangle\simeq 6 B_2 \frac{\alpha\omega_T^2}{\beta_0 t_0^{-1}}
\tilde{t}^{{2\beta_0}/{\alpha t_0\omega_m^2}}$. Comparison with
Eq. (\ref{MSD-over}) leads to the identifications $t_0\sim\beta_0^3\omega_m^{-4}$
and $H=1-\beta_0/{\alpha t_0 \omega_m^2}$. This expression
for $H$ constitutes the central result of this work. It shows that
the so-called Hurst exponent is a consequence of hydrodynamic and
elastic interactions between the particle and the viscoelastic medium.

This model explains the behavior of the MSD in the case of
constrained diffusion \cite{finite}. Fig. {\bf
\ref{FigCromosomas}} shows a comparison between theory (solid
line) and experiments of constrained diffusion of chromatin in
living cells \cite{cromosomas}.  The dashed line corresponds to a
similar model in the Markovian case. As expected, the short time
behavior of the MSD is subdiffusive. These effects are also
present in Fig. {\bf \ref {CompWirtz1}} (dotted lines). At short
times the behavior of MSD follows a power law on the scaled time
$\tilde{t}$ and saturates at long times, where confinement becomes
significant.  A more detailed model is reported in \cite{finite}.
\begin{figure}[tbp]
%\begin{center}
\includegraphics[width=0.6\linewidth]{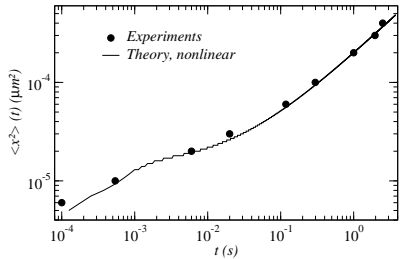}
%Comp-Fig6-WirtzBN.eps}
%\end{center}
\caption{MSD vrs time. Comparison between experiments (circles) and
theory (solid line) for the case of actin filament networks. The data
were taken from Ref. \cite{wirtz}. The theoretical model corresponds
to Eq. (\ref{MSD-nonlinear}) with $\langle x^2 \rangle_0\sim 4.4\cdot 10^{-6}\mu m^2$,
$D_0 \sim 9.3 \cdot 10^{-5}\mu m^2s^{-1}$,
$\beta_0\sim5.5\cdot 10^{9}s^{-1}$, $F_0\sim 3.8\cdot10^{-4} \mu m\,s^{-2}$, $t_0 \sim 0.1 s$,
$\lambda \sim 1.5\cdot 10^{-3} \mu m$, $B_2\sim 60$,
$\alpha\sim 1$.}
\end{figure}

A nonlinear model can also be discussed in terms of Eq.
(\ref{Smol-MNET}). If we assume that the force exerted on the
particle by the medium is of the form $\cos\left[\lambda^{-1}
\langle x\rangle(\tau)\right]$, with $\lambda$ the average
distance between the particle and the local boundaries, it can be
shown that \cite{finite}
\begin{equation}
\langle x^2 \rangle(t) \simeq \langle x^2 \rangle_0+ 2 D_0
\tau(t) + 4 \lambda^2 \tanh\left[\frac{F_0 D_0}{2 \lambda k_B
T}\tau(t)\right]^2, \label{MSD-nonlinear}
\end{equation}
where $\langle x^2 \rangle_0$ is some initial value. The behavior
of the MSD (\ref{MSD-nonlinear}) as a function of
time (solid line) is shown in Figs. \textbf{1} and \textbf{3}, and
compared with experiments (symbols). The plateau
is a signature of the existence of cage effects, which in our
model are related with the maximum value
of the elastic force ${\bf X}$. The agreement between theory and
experiments is good.

For Fig. {\bf 1} in the case with label $164$, the
values of the parameters are:
$\langle x^2 \rangle_0\sim 5\cdot 10^{-7}\mu m^2$,
$D_0 \sim 1.3 \cdot 10^{-4}\mu m^2s^{-1}$,
$\beta_0\sim5.5\cdot 10^{9}s^{-1}$,
$F_0\sim 1\cdot10^{-3} \mu m\,s^{-2}$, $t_0 \sim 0.1 s$,
$\lambda \sim 1.1\cdot 10^{-2} \mu m$, $B_2\sim 7.2$,
$\alpha\sim 1$ and we have defined  $\omega_m^2\sim F_0\lambda^{-1}$.
Finally, it is interesting to notice that the effective mobility
$\gamma(t)$ may be interpreted phenomenologically as the inverse Laplace
transform: ${\cal L}^{-1}\left[\omega/G''(\omega)\right]$. Taking
into account the relation between the MSD and the compliance $J$:
$\langle x^2 \rangle (t)\sim J(t)$, \cite{wirtz}, one obtains
$\gamma(t)={\cal L}^{-1}\left[\omega^2 J(\omega)\right]$.

\section{Conclusions}

By following two different approaches, in this article we have
described the anomalous diffusion of a Brownian particle in
microrheological experiments. In the linear force case, we have
shown that the generalized Langevin description with fractionary
Gaussian noise explains well the behavior of the microrheological
properties of the viscoelastic medium. We have also shown the
equivalence of this description with generalized Fokker-Planck
equations having time dependent coefficients. The comparison with
a thermokinetic formalism gives a plausible justification for the
physical nature of the exponent $H=1-\beta_0/\alpha t_0
\omega_m^2$ used in the GLE description. The thermokinetic
formalism constitutes a powerful generalization of the theory
allowing the formulation of non-Markovian models even in the case
of non-linear forces. The agreement between theory and experiment
is good.

%\begin{large}
\emph{Acknowledgments.}
%\end{large}
I am indebted to Prof. J. Miguel Rub\'i for illuminating
discussions concerning this work. Thanks go also to Profs. R. F.
Rodr\'iguez and A. Gadomski and to Drs. A. P\'erez-Madrid, M.
Mayorga and L. Romero-Salazar. This work was supported by
UNAM-DGAPA under the grant IN-108006.


\begin{thebibliography}{1}

\bibitem{trimper} S. Trimper, K. Zabrocki, M. Schulz, Phys. Rev. E
\textbf{70}, 056133-1-056133-7 (2004).

\bibitem{mason2000} T. G. Mason, Rheol. Acta \textbf{39},
371-378 (2000).

\bibitem{wirtz} J. Xu, V. Viasnoff, D. Wirtz Rheol. Acta \textbf{37},
387-398 (1998).

\bibitem{finite} I. Santamar\'ia-Holek, J. M. Rub\'i, J. Chem. Phys.
\textbf{125}, 064907-1-064907-4 (2006); I. Santamar\'ia-Holek, J.
M. Rub\'i, A. Gadomski, J. Phys. Chem. B \textbf{111}, 2293-2298
(2007).

\bibitem{cromosomas} W. F. Marshall, \textit{et al.}, Current Biology
\textbf{7}, 930-939 (1997).

\bibitem{revMNET} D. Reguera, J. M. G. Vilar, J. M. Rub\'{i}, J. Phys. Chem. B \textbf{109}, 21502-21515 (2005).

\bibitem{adelman} S. A. Adelman, J. Chem. Phys. \textbf{64}, 124-130 (1976).

\bibitem{tokuyama} M. Tokuyama and H. Mori, Prog. Theor. Phys. \textbf{55},
411-429 (1976).

\bibitem{libroZ} R. Zwanzig, \textit{Nonequilibrium Statistical Mechanics }, Oxford University Press, New York, 2001.

\bibitem{donor} S. C. Kou, X. S. Xie, Phys. Rev. Lett. \textbf{93}, 180603-1-180603-4 (2004).

\bibitem{balescu} R. Balescu \textit{Statistical Dynamics}, Imperial College Press, Singapore, 1997.

\bibitem{saxena} A. M. Mathai, R. K. Saxena, H. J. Haubold,
Astrophys. Space Sci. \textbf{305}, 283-288
 (2006).

\bibitem{PHYSA03} I. Santamar\'{\i}a-Holek and J. M. Rub\'{\i}, Physica A \textbf{326},
384-389 (2003).

\bibitem{oxtoby} S. Okuyama and D. W. Oxtoby, J.Chem. Phys. \textbf{84}, 5824-5829 (1986).

\bibitem{groot} S. R de Groot and P. Mazur, \textit{Non-Equilibrium
Thermodynamics}, Dover, New York, 1984.

\bibitem{Adam07} A. Gadomski, Physica A {\bf 373}, 43-57 (2007).

\bibitem{brenner} J. Happel, H. Brenner, \textit{Low Reynolds number
hydrodynamics}, Kluwer, Dodrecht, 1991.

\bibitem{saarlos} C. W. J. Beenakker, W. van Saarloos, and P. Mazur, Physica A
\textbf{127}, 451-472 (1984).

\bibitem{Agus05} A. P\'{e}rez-Madrid, J. Chem. Phys. \textbf{122},
214914-1-214914-6 (2005).

\bibitem{signpost} I. Santamar\'ia-Holek, J. M. Rub\'i, "Anomalous Diffusion in
Intracellular Transport" in \emph{Physics of Complex Systems and
Life Sciences}, edited by A. Wagemakers and M. A. F. Sanjuan,
Research Signpost, Kerala, 2008.



\end{thebibliography}
\end {document}